\documentstyle[epsfig,emulateapj]{article}

\lefthead{Meier}
\righthead{THICK DISK-JET ASSOCIATION}

\received{ }

\slugcomment{Submitted to The Astrophysical Journal Letters.}

\begin{document}

\hyphenation{ }

\title{The Association of Jet Production with Geometrically Thick \\
    Accretion Flows and Black Hole Rotation}

\author{D. L. Meier}
\affil{Jet Propulsion Laboratory, California Institute of Technology,
    Pasadena, CA 91109}

\begin{abstract}
A model is presented in which the strongest radio-emitting 
jet outflows are produced in black hole systems when the accretion 
is a geometrically thick ($H/R \sim 1$) inflow ({\it e.g.}, ADAF, CDAF) {\em 
and} if the black hole is rotating.  For galactic black hole candidates, 
the model naturally accounts for 
the observed correlation of jet outflow with the black hole hard emission 
state and predicts an association of strong jets with rapid black hole rotation.
When extended to the supermassive case, the model accounts for 
the highest radio galaxy and quasar jet powers and provides additional 
theoretical support for the ``spin paradigm'', which asserts that radio 
loud quasars are produced by Kerr holes and radio quiet ones by Schwarzschild 
holes.  In some cases, the angular momentum and energy outflow in the 
jet may be large enough to significantly alter the structure of 
the accretion flow from that predicted by current models.
\end{abstract}


\section{Introduction}
There is now substantial evidence that, in binary black hole candidate 
X-ray sources (BHCs), a radio-emitting, jet outflow is produced only 
when the X-rays are in the hard state.  That is, the jet is ``quenched''
when the dominant accretion disk emission is from soft (2-10 keV)
photons and little or no emission is seen in the hard (20-100 keV) region.
Sources which display this behavior are GX 339-4 (\cite{f99a}; \cite{f99b}), 
GRO J1655-40 (at least in the first few 1994 outbursts; \cite{h95}), 
GRS 1915+105 (\cite{h97}; \cite{f99a}), Cyg X-3 (\cite{mc99}), and 
Cyg X-1 (\cite{br99}).

In addition, there is some evidence that those black hole systems which 
produce a strong jet contain Kerr (rapidly-rotating) black holes, although 
this result requires additional confirmation.  Recent 
attempts to measure the angular momentum parameter $j \equiv J/(GM^2/c)$
({\it i.e.}, ``$a/M$'' in geometric units) indicate large values in the 
superluminal sources GRS 1915+105 and GRO J1655-40 ($j=0.95$) and smaller 
values in the non-superluminal systems Cyg X-1 and GS 1124-68 (\cite{cui98}).  
Furthermore, in the supermassive black hole case this 
``spin paradigm'' has useful advantages for explaining the difference 
between radio loud quasi-stellar radio sources (QSRs) and radio quiet 
quasi-stellar objects (QSOs) and why giant radio sources occur only 
in elliptical galaxies (\cite{wc95}; \cite{m99}).

This letter shows that these properties are a natural consequence of black 
hole accretion and can be explained by combining three 
current areas of astrophysical theory into a single model:  1) modern disk 
accretion theory, including geometrically thick inflows at low accretion rate, 
standard disks at intermediate rate, and thermally unstable disks at high 
rates;  2) standard jet-production theory, in which the jet power 
is a strong function of the poloidal magnetic field strength and of the 
angular velocity of that field; and 3) dynamo theory for the accretion 
viscosity, in which the poloidal field strength is a strong function of 
the disk height.  Thin disk flows, despite their rapid Keplerian rotation, 
do not have strong vertical fields and, therefore, should not produce strong jets.
Geometrically thick accretion flows, on the other hand, are expected to 
have strong vertical fields, but suffer from slower rotation.  Only 
when the disk is thick {\em and} the geometry also rotates rapidly, as it 
does near a Kerr hole, do we obtain both strong fields and rapid rotation 
and, therefore, strong jets.

\section{Supermassive Black Holes --- Radio Galaxies and Quasars}

\label{smbh_section}

\subsection{Maximum Total Observed and Predicted Jet Power}

The maximum observed total jet power possible from supermassive black hole 
systems will be used to constrain the models for jet production.  To do so, 
the upper envelope of the OLWL diagram (\cite{ol89};  \cite{ow91}; \cite{lo96})
for powerful radio sources is used to give a maximum radio power as a 
function of the host galaxy's red optical magnitude.  This relation is given 
approximately by
\begin{displaymath}
{\rm log} P_{rad}^{max}  \; \approx  \; -0.56 M_R \, + \, 14.9
\end{displaymath}
with the radio power $P$ in ${\rm W \, Hz^{-1}}$.  Secondly, the Kormendy \& 
Richstone correlation for red magnitude (\cite{kr95}) 
\begin{displaymath}
{\rm log} \, m_9 \; \approx \; -0.5 M_R \, - 11.3
\end{displaymath}
is used to determine radio power as a function of black hole mass, 
where $m_9 = M_{\bullet} / 10^9  M_{\sun}$.  
Finally, a factor of $\sim 10^{-11} - 10^{-12} \, {\rm Hz}^{-1}$ (\cite{bick95}), 
corresponding to a fraction $\epsilon \sim 10^{-2} - 10^{-3}$ of the jet power emitted in 
a $10^9 \, {\rm Hz}$ bandwidth, is used 
to convert radio power to total jet power, yielding the following expression
for an estimate of maximum {\em total} jet power
\begin{equation}
\label{obs_max}
L_{jet}^{max} \; = \; 10^{46} \, {\rm erg \, s^{-1}} \, m_9^{1.12} \, \epsilon_{-2.5}^{-1}
\end{equation}
Any mechanism of jet production must be able to achieve this maximum power, 
not only instantaneously but also sustained over long periods of time, as 
well as scale roughly linearly with black hole mass.

To compare this observed power with theoretical estimates, jets from both 
standard disks and advection-dominated accretion flows (ADAFs) will be 
considered, comparing the results for both Schwarzschild and Kerr black holes.  
This discussion is similar to that of (\cite{lop99}; LOP), except that 
the jet power from non-spinning and spinning black hole systems will be 
contrasted, rather than comparing the power from different components (disk 
and hole) within a single system; and geometrically thick accretion flows 
will be considered as well as thin disks.  
As LOP point out, when the hole is spinning, the power generated by the 
magnetic field threading the disk (the Blandford-Payne process; \cite{bp82}) 
and that from the field threading the black hole (the Blandford-Znajek 
process; \cite{bz77}) will be comparable, at best.  
However, as pointed out earlier (\cite{m99}), even if the BZ process is 
neglected entirely, the jet power contributed by the field threading the disk 
alone still will be a function of the black hole spin, since the rotating 
metric contributes to the rotation of the magnetic field.
That is, as viewed from an outside observer at infinity in the Boyer-Lindquist 
reference frame, the disk angular velocity is a sum of its angular 
velocity relative to the local metric $\Omega '$ ({\it i.e.}, in the Zero 
Angular Momentum Observer or ZAMO frame) plus the angular velocity of the 
metric itself in the Boyer-Lindquist frame $\omega \equiv - g_{\phi t} / 
g_{\phi \phi}$
\begin{displaymath}
\Omega \; = \; \Omega ' \, + \, \omega
\end{displaymath}

The structure equations for accretion flows ({\it e.g.}, \cite{ss73}; 
\cite{nt73}; \cite{ny95}) present a solution for the disk 
magnetic field which generally is identified with the dominant azimuthal 
component $B_{\phi}$.  However, models and simulations of jet production 
(\cite{bz77}; \cite{bp82}; \cite{megpl97}) 
show that it is the strength of the {\em poloidal} field component ($B_p 
= [B_R^2 + B_Z^2]^{1/2}$) that is largely responsible for the production of 
a jet,  with the output power given by 
\begin{equation}
\label{ljet_basic}
L_{jet} \; = \; B_{p0}^2 \, R_{0}^4 \, \Omega_{0}^2 \, / \, 4c
\end{equation}
where $R_{0}$ is the characteristic size of the jet-formation region, and 
$B_{p0}$ and $\Omega_{0}$ are, respectively, the poloidal field and angular velocity 
of the field in that region as measured by an external observer.  Now, 
LOP have argued, and dynamo simulations tentatively have confirmed (\cite{k99}), 
that the strength of the poloidal magnetic field is limited by the 
vertical extent of the turbulent eddies.  That is, $B_{p0} / B_{\phi 0} 
\approx (H_{0} / R_{0})^n$, where $H_{0}$ is the vertical half-thickness 
of the disk at radius $R_{0}$, and the exponent $n \sim 1$, yielding 
an estimate of the jet power of 
\begin{equation}
\label{ljet_with_h}
L_{jet} \; = \; B_{\phi 0}^2 \, H_{0}^2 \, R_{0}^2 \, \Omega_{0}^2 \, / \, 4c
\end{equation}
which can be used to calculate the instantaneous jet power once 
$B_{\phi 0}$, $H_{0}$, and $\Omega_{0}$ are determined using a particular 
disk model.

\subsection{Jet Power from Standard Thin Accretion Disks}

First, standard disks will be considered.  For illustrative purposes, 
the ``middle region'' solution will be used, where gas pressure and 
electron-scattering opacity are dominant.  
For the Schwarzschild case ($\omega = 0$), a characteristic 
radius of $R_{0} = 7 GM/c^2$ (just outside the last stable orbit) will be 
used, along with the Keplerian angular velocity of $\Omega_{0} ' = 7^{-3/2} 
(GM/c^3)^{-1}$.  For the Kerr case, since the disk extends inside the 
ergosphere (inside $R = 2 GM/c^2$),  a characteristic radius of $1.5 GM/c^2$ 
will be used, along with $\omega_{0} = \omega(R_{0}) \approx 0.6 (GM/c^3)^{-1}$.  
This yields
\begin{eqnarray}
\label{schw_ss}
L_{jet}^{Schw-SS} & \approx & 10^{42.6} \, {\rm erg \, s^{-1}} \, 
(\alpha_{-2}^{SS})^{-1/10} \, m_9^{9/10} \, \dot{m}_{-1}^{6/5} ~~~~~
\\
\nonumber
L_{jet}^{Kerr-SS} & \approx & 10^{43.5} \, {\rm erg \, s^{-1}} \, 
(\alpha_{-2}^{SS})^{-1/10} \, m_9^{9/10} \, \dot{m}_{-1}^{6/5}  ~~~~~
\\
\label{kerr_ss}
& & ~~ (1 + 1.1j + 0.29 j^2)
\end{eqnarray}
where $\alpha_{-2}^{SS}$ is the viscosity parameter 
for standard thin disks in units of $0.01$, 
$\dot{m}_{-1} \equiv \dot{M} / 0.1 \dot{M}_{Edd}$ is the accretion 
rate scaled to that producing one-tenth of an Eddington luminosity, and 
the Eddington accretion rate is defined as $\dot{M}_{Edd} = 22 M_{\sun} \, 
{\rm yr^{-1}} \, m_9$ (assuming $10 \%$ accretion efficiency).  One-tenth 
the Eddington accretion rate is used as the 
scale value since it is roughly the rate at which the low/hard to high/soft 
black hole state transition is believed to occur (\cite{nmq98}).

As LOP point out, the black hole spin makes little difference for standard 
thin disks.  Within the Kerr system, it contributes only about a third of 
the power.  And the ratio of power from a Kerr to a Schwarzschild hole 
is $\sim 20$, with most of this due to the deeper potential well in the Kerr 
case.  Furthermore, the jet power for a $10^9 M_{\sun}$ hole is limited 
to $\sim 7 \times 10^{43} \, {\rm erg \, s^{-1}}$ --- two orders of magnitude 
below our target in equation (\ref{obs_max}).  So, while Shakura \& Sunyaev 
disks might account for some of the weakest radio sources (in particular, 
the ``radio quiet'' QSOs, when there is significant optical emission 
from the disk), they cannot account for the most powerful radio galaxies 
and quasars.

\subsection{Jet Power from Advection-Dominated Accretion Flows}

Next, ADAFs will be considered, as an example of a geometrically thick 
accretion flow.  Note that this case will include 
coronae of thin disks in which a dominant fraction of the gravitational 
binding energy of the accreting material is dissipated in the corona 
rather than in the disk.  For these flows $H_{0} \sim R_{0}$, so 
$B_{p0} \sim B_{\phi 0}$.  With the ADAF solutions (\cite{nmq98}), 
both the Schwarzschild and Kerr cases yield
\begin{eqnarray}
\label{schw_adaf}
L_{jet}^{Schw-AD} & \approx & 10^{44.7} \, {\rm erg \, s^{-1}} \, 
(\alpha_{-1/2}^{AD})^{-1} \, m_9 \, \dot{m}_{-1} \, f^2 ~~~~~
\\
\nonumber
L_{jet}^{Kerr-AD} & \approx & 10^{46.0} \, {\rm erg \, s^{-1}} \, 
(\alpha_{-1/2}^{AD})^{-1} \, m_9 \, \dot{m}_{-1} \, g^2 ~~~~~
\\
\label{kerr_adaf}
& & ~~ (0.55 f^2 + 1.5 f j + j^2)
\end{eqnarray}
where $f \equiv \frac{\Omega '_{0}}{\Omega '_{0 N-Y}}$ is the ratio of the 
actual angular velocity to that calculated by Narayan and Yi, and $g \equiv 
\frac{B_{\phi 0}}{B_{\phi N-Y}}$ is a similar ratio for the azimuthal 
magnetic field.  Note that, as is customary, the viscosity parameter 
for ADAFs is now scaled to $\sim 0.3$ instead of $0.01$.  While considerably 
larger than the jet powers calculated for thin disks, these expressions 
are not simply $(R_{0}/H_{0})^2$ times equations (\ref{schw_ss}) and 
(\ref{kerr_ss}).  This is because ADAF azimuthal fields are weaker 
than in thin disks, and the rotation rate of the inflowing plasma 
is somewhat smaller (0.4 Keplerian, as determined by Narayan \& Yi). 
In fact, $\Omega_{0} '$ is very sensitive to the exact value of the 
internal adiabatic index:  $f$ could be considerably less than unity 
($\Gamma = 5/3$ yields $f=0$!), so the results have been scaled to $f^2$.
Equation (\ref{schw_adaf}), therefore, is probably an upper limit for the 
jet power from an advection-dominated flow into a Schwarzschild hole.

On the other hand, in the Kerr case, the actual values of $B_{\phi 0}$ may 
be {\em higher} than that calculated by Narayan \& Yi, since these authors did 
not take the field-enhancing shear in the Kerr metric into account (see 
{\it e.g.}, \cite{m99}). It is not unreasonable to expect an enhancement of 
at least $g \sim \omega(R_{0}) / \Omega_{0 N-Y} \sim 2.3$, with a 
corresponding factor of 5 or more increase in jet power.  This alone would 
put the Kerr-ADAF jet power for a $10 \%$ Eddington accretion rate well above 
the target of $10^{46} \, {\rm erg \, s^{-1}}$.  Equation (\ref{kerr_adaf}), 
therefore, is likely to be a {\em lower} limit on the jet power from an 
advection-dominated flow into a Kerr hole.  Unlike the thin disk case, 
then, the black hole spin in the ADAF case is very important, especially 
if $f \rightarrow 0$.  The ratio of Kerr to Schwarzschild jet power 
($\sim 20 g^2/f^2$) is potentially 100 or more.

The jet power in equation (\ref{kerr_adaf}) is a substantial fraction 
of the accretion luminosity
\begin{equation}
\label{kerr_adaf_ratio}
\frac{L_{jet}^{Kerr-AD}}{L_{acc}} \; = \; 0.80 \, (\alpha_{-1/2}^{AD})^{-1} 
\, (0.14 f^2 + 0.74 f j + j^2) \, g^2
\end{equation}
Consequently, the angular momentum and energy carried away from the ADAF 
will have a substantial effect on the structure of the inflow, which 
must be taken into account in the ADAF solutions.  (This is in addition to 
any thermal wind that may be part of an Advection-Dominated Inflow/Outflow 
Solution or ADIOS which also must be taken into account. See, {\it e.g.}, 
\cite{bb99}.)  Indeed, the ratio could easily 
exceed unity, removing {\em more} angular momentum and rotational energy from 
the inflow than it has in the ZAMO frame.  This could reverse the spin 
of the ADAF in the ZAMO frame, giving it {\em negative} angular momentum, 
and ultimately spinning down the hole slightly when the accreting matter 
finally enters the horizon.  Such a scenario is quite plausible in Kerr 
geometry and, indeed, has been seen in magnetohydrodynamic (MHD) numerical 
simulations of jet production near rotating holes (\cite{koide00}). 
This process is a type of Penrose mechanism, with the disk particles 
scattered into negative angular momentum orbits by a torsional MHD 
wave (the jet) that, in turn, is ``scattered'' upward and outward to infinity.  
Similar in character to the Punsly-Coroniti ``ergospheric winds''
(\cite{pc90}), this process can extract angular momentum and energy from 
the black hole even without the magnetic field threading the horizon.  

To summarize, the most powerful extragalactic jet sources require 
both a geometrically thick inflow {\em and} a rapidly spinning black 
hole to reach the observed maximum jet powers.

\section{Stellar Mass Black Holes --- Microquasars}

The possible association of black holes having potentially high spin 
with binary X-ray sources that produce jets, and the association of 
the onset of jet production in these systems with a hard emission 
state, is very similar in character to the need for both spin and 
advective accretion flow to produce extragalactic jets discussed above.
In fact, diagrams drawn depicting this association (\cite{f99a}) can 
be qualitatively, and to a great extent quantitatively, 
understood using the above equations.  
For the disk structure in different accretion regimes, the discussion 
below will adopt the solutions of \cite{ch95} and \cite{esin98}.  Equations 
(\ref{schw_ss}) - (\ref{kerr_adaf}) will be used to calculate the jet 
powers, with the appropriate scaling of mass and accretion rate.  Note 
that the jet power predicted by these equations is 
primarily kinetic energy and Poynting flux; the actual 
radio luminosity emitted will be much less.  For example, during the 
first 1994 outburst in GRO J1655-40, which was thought to have reached 
accretion rates approaching $\dot{m} \sim 1$, the total jet power is 
$\sim 10^{38-39} \, {\rm erg \, s^{-1}}$ (\cite{m96}), while the radio 
luminosity is only $\sim 10^{32} \, {\rm erg \, s^{-1}}$ (\cite{h95}).  

For low accretion rates, $\dot{m} < \dot{m}_{A} \approx 0.1 \, 
(\alpha_{-1/2}^{AD})^2$ accreting black holes are in the low/hard 
state --- steady accretion with a substantial fraction of the X-rays 
in the 20-100 keV band.  This corresponds to an ADAF solution, which 
is favored over a thin disk solution in this regime.  
As the ADAF is stable, and expected to produce a jet power of up to 
$\sim 5 \times 10^{38} \, {\rm erg \, s^{-1}} \, m_1 \, 
\dot{m}_{-1} \, j^2$ (equation \ref{kerr_adaf}), a continuous jet outflow 
that is roughly proportional to the accretion rate is expected.  This is, 
in fact, seen in several microquasars (\cite{f99b}).  Note that, although 
proportional to the accretion rate, the power in this model comes mostly 
from the rotation of the black hole and not from the accretion energy of 
the infalling matter. However, the efficiency of extracting the rotational 
energy of the hole {\em is} proportional to $\dot{m}$.

For $\dot{m} > \dot{m}_A$ the disk enters the high/soft state (\cite{nmq98}), in 
which the hot ADAF largely disappears and the disk takes on an optically 
thick, cooler structure similar in nature to the standard (\cite{ss73}) 
thin disk solutions.  Here the cool disk is modeled with the 
Shakura \& Sunyaev ``middle'' region.  
The expected jet power from such a thin disk with a weak 
poloidal field (equation \ref{kerr_ss}) is $\sim 5 \times 10^{36} \, 
{\rm erg \, s^{-1}} \, m_1^{9/10} \, \dot{m}_{-1}^{6/5}$ --- a hundred 
times weaker than in the ADAF case. This compares well to the jet suppression 
of $> 35$ seen in GX 339-4 (\cite{f99a}) when it goes into the 
soft state, and is some of the best evidence that a strong jet cannot 
be produced unless the accretion flow is geometrically thick and hot.

For high accretion rates ($\dot{m} \gtrsim 0.5-1.0$) the disk enters 
the ``very high state'', in which the disk is very unstable.  In GRS 
1915+105, for example, the disk goes through cycles of $\sim 20$ minutes 
in duration, which begin with a dramatic dip in the soft X-rays, followed 
by an increase in the hard X-ray emission for several minutes.  Just 
as the hard emission abruptly disappears, a radio jet is ejected, followed 
by a softening of the spectrum again.  A possible explanation for 
this behavior is the onset of a thermal ``limit cycle'', in which the 
disk oscillates between two unstable accretion states (standard soft 
and hard ADAF).  Detailed modeling of this case by \cite{sm98} 
suggests that the disk first bloats to a time-dependent, hot, 
gas-pressure-dominated ADAF-like state; then it drains rapidly into 
the hole; and finally a thin radiation-pressure-dominated thin disk 
re-fills, only to become unstable and bloat again.  For a 
$10 \, M_{\sun}$ hole, the period of the limit cycle is on the order 
of $\sim 1000$ seconds, with the active ADAF phase lasting only 
$\sim 1\%$ of that time.  

There is some uncertainty in this model, however.  Standard disk 
stability theory (\cite{ss76}) predicts that unstable behavior 
should occur above the rather low accretion rate of $\dot{m} \gtrsim 
0.02 (\alpha_{-2}^{SS} \, m_1)^{-1/8}$, when the innermost part of 
the disk becomes radiation pressure dominated.  Indeed, the unstable 
disk of \cite{sm98} 
had $\dot{m} = 0.06$.  This rate is so low, however, that it predicts 
that all disks with $\dot{m} > \dot{m}_A$ should be unstable; that is, 
there should be no stable high/soft state even in the case of X-ray 
binaries; at most the disk should be in a cool state for a portion of 
the limit cycle (20 minutes).  This is in direct conflict with the 
observations:  the disk stays in the high/soft state often for months.
Using a more careful analysis, \cite{ch95} derive 
a modified transition accretion rate of
\begin{equation}
\label{mdot_r}
\dot{m}_R \; = \; 0.3 \, (\alpha_{-2}^{SS} \, m_1)^{-1/8}
\end{equation}
(minimum of the solid lines in their figure 2), above which limit cycle 
behavior should set in.  Equation (\ref{mdot_r}) agrees well with the 
observations of GRS 1915+105 and GX 339-4.  Note also that the solutions 
of \cite{ch95} also predict that unstable behavior should continue for 
accretion rates well in excess of the Eddington rate ($\dot{m} \lesssim 
20$ or so), above which a steady, radiation-pressure-dominated ADAF 
should set in.

Ascribing a high accretion rate ($\dot{m} \sim 1$) to 
the hard temporary ADAF state, equation (\ref{kerr_adaf}) predicts 
a {\em short-term} jet power during the very high state of $\sim 6 \times 10^{39} \, 
{\rm erg \, s^{-1}} \, m_{3/2} \, \dot{m} \, g^2$ for a rapidly-rotating 
black hole.  Again, allowing for possible metric-shear enhancement 
of the magnetic field, instantaneous jet powers well in excess of 
$10^{40} \, {\rm erg \, s^{-1}}$ in GRS 1915+105 are possible in this model.

\section{A Grand Unified Scheme for Active Galactic Nuclei}

Finally, we turn once again to the supermassive case.  The above concepts 
allow us to begin to build a scheme that unifies radio loud and radio quiet 
AGN.  It is similar to that discussed earlier (\cite{m99}), but with the addition of 
geometrically thick advection-dominated flows.  

Since radiation pressure is even more dominant in supermassive black 
hole disks than in stellar mass ones ($\dot{m}_{R} \propto m_1^{-1/8}$), 
it is likely that $\dot{m}_{R}$ will be less than $\dot{m}_{A}$ for 
some large hole mass (from equation [\ref{mdot_r}], for $M_{\bullet} \gtrsim 7 \times 10^{4} \, 
M_{\sun}$).  That is, for supermassive holes, a stable high/soft state 
should not exist. Unlike before in the stellar mass case, the possible 
lack of a high/soft state is {\em not} 
in conflict with observations.  A quasar could be in the very high state 
at the present time and show no signs of limit cycle behavior, as the time 
scale for the cycle should be $10^8$ times longer than the stellar-mass black 
hole case (several thousand years).  Only every few thousand years would a 
quasar enter a very active phase that lasted a few decades.  During that time 
a temporarily high accretion rate ($\dot{m} \sim 1$) period may occur, with 
jet powers also given by equations (\ref{schw_adaf}) and (\ref{kerr_adaf}), 
while during the quiescent phase the jet powers would be given by equations 
(\ref{schw_ss}) and (\ref{kerr_ss}).  The final jet power on the kiloparsec 
scale (integrated over thousands of years and several limit cycles) would be a 
sum of the two pertinent equations, weighted by the duty cycle of each phase
--- that is, approximately equation (\ref{schw_adaf}) or (\ref{kerr_adaf}), 
with $\dot{m}$ replaced by the accretion rate averaged over the long term.

Accreting supermassive black holes, then, should have only two basic 
states for each of the Schwarzschild and Kerr cases:  low/hard (ADAF) or 
very high (geometrically thin, radiation-pressure-dominated and unstable 
flows).  We consider the very supermassive ($10^{7-9} M_{\sun}$/quasar) 
case first.  The Schwarzschild low/hard case is described only by equation 
(\ref{schw_adaf}), and corresponds to weak radio cores with little or no line 
emission (``Class B'' according to \cite{jw99}).  As discussed 
above, the Schwarzschild {\em very high} disk case is a linear combination of 
equations (\ref{schw_ss}) and (\ref{schw_adaf}), yielding a long-term jet 
power of 
$L_{jet}^{Schw} \; \sim \; 5 \times 10^{43} \, {\rm erg \, s^{-1}} \, f^2$
for an ADAF duty cycle of $1\%$ and a $10^9 M_{\sun}$ hole.  In the latter 
state the hole usually would produce significantly more optical-UV 
radiation than in the low/hard state, appearing as a radio quiet 
``Class A'' object (QSO).  
The Kerr low/hard case is described by equation (\ref{kerr_adaf}), and 
corresponds to a Class B radio galaxy (either \cite{fr74} [FR] type I 
or type II without strong nuclear emission lines).  The Kerr very high 
state is again a linear combination of equations (\ref{kerr_ss}) and 
(\ref{kerr_adaf}), yielding $L_{jet}^{Kerr} \; \sim \; 
10^{45} \, {\rm erg \, s^{-1}} \, g^2$ for a $10^9 M_{\sun}$ hole and a $1\%$
duty cycle --- a radio loud Class A object (quasar or FR II BLRG) with a jet 
perhaps hundreds of times more powerful than in the above 
radio quiet case.  

In the moderately supermassive ($10^{5-7} M_{\sun}$/Seyfert) case, 
significant optical emission should be present in both the low/hard and 
very high states, as that emission would be produced in the middle 
regions of the disk, which are not affected by state transitions.  In the 
very high state, during the thin-disk portion of the limit cycle, the peak 
emission would be in the extreme ultraviolet ($\sim 0.1$ keV).  Such 
objects should appear similar to the highly-variable, ultrasoft narrow-line 
Seyfert 1 galaxies (\cite{b00}), whereas in the low/hard emission state they 
should appear more like the classical hard X-ray selected Seyferts. 
Their radio jet properties should be similar to those of the quasars above, 
though $100$ times weaker, with the ultrasoft objects producing a jet 
only when they enter the thick-disk portion of their limit cycle.

\acknowledgments

The author is pleased to acknowledge the Aspen Center for Physics and workshops
there on Astrophysical Dynamos and on Acceleration and Collimation of 
Astrophysical Jets.
He is especially grateful for discussions with R. Antonucci, R. Fender, 
R. Hjellming, S. Koide, and E. Vishniac. Bob Hjellming was a good friend 
to many and will be greatly missed.
This research was carried out at the Jet Propulsion Laboratory, California 
Institute of Technology, under contract to 
NASA.


\begin{thebibliography}{}

\bibitem[Blandford \& Begelman 1999]{bb99}
    Blandford, R. D. \& Begelman, M. C. 1999, \mnras, 303, L1. 

\bibitem[Bicknell 1995]{bick95}
    Bicknell, G. V. 1995, \apjs, 101, 29. 

\bibitem[Blandford \& Znajek 1977]{bz77}
    Blandford, R. D. \& Znajek, R. 1977, \mnras, 179, 433 (BZ). 

\bibitem[Blandford \& Payne 1982]{bp82}
    Blandford, R. D. \& Payne, D. G. 1982, \mnras, 199, 883 (BP). 

\bibitem[Brandt 2000]{b00}
    Brandt, W. N. 2000, in {\it Highly Energetic Physical Processes and 
    Mechanisms for Emission from Astrophysical Plasmas}, IAU Symp. 195, 
    p. 207.

\bibitem[Brocksopp {\it et al.} 1999]{br99}
    Brocksopp, C. {\it et al.} 1999, \mnras, 309, 1063.

\bibitem[Chen {\it et al.} (1995)]{ch95}
    Chen, X., Abramowicz, M.A., Lasota, J.-P., Narayan, R., \& 
    Yi, I. 1995, \apj, 443, L61.

\bibitem[Cui {\it et al.} 1998]{cui98}
    Cui, W. Zhang, S.N., \& Chen, W. 1998, \apj, 492, L53.

\bibitem[Esin {\it et al.} (1998)]{esin98}
    Esin, A.A., Narayan, R., Cui, W., Grove, J.E., \& Zhang, S.-N. 1998, 
    \apj, 505, 854. 

\bibitem[Fanaroff \& Riley 1974]{fr74}
    Fanaroff, B. L. \& Riley, J. M. 1974, \mnras, 164, 31P. 

\bibitem[Fender {\it et al.} 1999]{f99a}
    Fender, R. {\it et al.} 1999, \apj, 519, L165-168.

\bibitem[Fender 1999]{f99b}
    Fender, R.P. 1999, in Black Holes in binaries and galactic nuclei, 
    in press.  Available as astro-ph/9911176.

\bibitem[Harmon {\it et al.} 1995]{h95}
    Harmon, B.A. {\it et al.} 1995, \nat, 374, 703.

\bibitem[Harmon {\it et al.} 1997]{h97}
    Harmon, B.A. {\it et al.} 1997, \apj, 477, L85.

\bibitem[Honma {\it al.} 1991]{hmk91}
    Honma, F., Matsumoto, R., \& Kato, S. 1991, \pasj, 43, 147.

\bibitem[Khanna 1999]{k99}
    Khanna, R. 1999, in {\it Stellar Dynamos: Nonlinearity and Chaotic Flows}, 
    ed. A. F. Moz \& M. Nunez, ASP Conf. Series, Vol. 178.

\bibitem[Koide et al. 2000]{koide00}
    Koide, S., Meier, D.L., Shibata, K., \& Kudoh, T. 2000, \apj, 
    in press. 

\bibitem[Jackson \& Wall 1999]{jw99}
    Jackson, C. A. \& Wall, J. V. 1999, \mnras, 304, 160. 

\bibitem[Kormendy \& Richstone 1995]{kr95}
    Kormendy, J. \& Richstone, D. 1995, Ann. Rev. Astr. Ap., 33, 581. 

\bibitem[Ledlow \& Owen 1996]{lo96}
    Ledlow, M. J. \& Owen, F. N. 1996, \aj, 112, 9. 

\bibitem[Livio, Ogilvie, \& Pringle 1999]{lop99}
    Livio, M., Ogilvie, G. I., \& Pringle, J. E. 1999, \apj, 512, 100 
    (LOP). 

\bibitem[McCollough {\it et al.} 1999]{mc99}
    McCollough, M.L., {\it et al.} 1999, \apj, 517, 951.

\bibitem[Meier 1996]{m96}
    Meier, D. L. 1996, \apj, 459, 185. 

\bibitem[Meier 1999]{m99}
    Meier, D. L. 1999, \apj, 522, 753. 

\bibitem[Meier {\it et al.} 1997]{megpl97}
    Meier, D.L., Edgington, S., Godon, P., Payne, D.G., \& 
    Lind, K.R. 1997, \nat, 388, 350.  

\bibitem[Narayan, Mahadevan, \& Quataert 1998]{nmq98}
    Narayan, R., Mahadevan, R., \& Quataert, E. 1998, in 
    {\it The Theory of Black Hole Accretion Disks}, eds. 
    Abramowicz, Bjornsson, \& Pringle (Cambridge: Cambridge Univ. 
    Press).

\bibitem[Narayan \& Yi 1995]{ny95}
    Narayan, R. \& Yi, I. 1995, \apj, 444, 231. 

\bibitem[Novikov \& Thorne 1973]{nt73}
    Novikov, I. D. \& Thorne, K. S. 1973, in Black Holes, eds. C. DeWitt 
    \& B. DeWitt, (New York: Gordon \& Breach), p. 343. 

\bibitem[Owen \& Laing 1989]{ol89}
    Owen, F. N. \& Laing, R. A. 1989, \mnras, 238, 357. 

\bibitem[Owen \& White 1991]{ow91}
    Owen, F. N. \& White, R. A. 1991, \mnras, 249, 164. 

\bibitem[Punsly \& Coroniti 1990]{pc90}
    Punsly, B. \& Coroniti, F. V. 1990, \apj, 354, 583. 

\bibitem[Shakura \& Sunyaev 1973]{ss73}
    Shakura, N. I. \& Sunyaev, R. A. 1973, \aap, 24, 337.  

\bibitem[Shakura \& Sunyaev 1976]{ss76}
    Shakura, N. I. \& Sunyaev, R. A. 1976, \mnras, 175, 613.  

\bibitem[Szuszkiewicz \& Miller (1998)]{sm98}
    Szuszkiewicz, E. \& Miller, J.C. 1998, \mnras, 298, 888. 

\bibitem[Wilson \& Colbert 1995]{wc95}
    Wilson, A. S. \& Colbert, E. J. M. 1995, \apj, 438, 62. 


\end{thebibliography}
\end{document}